\documentclass[conference]{IEEEtran}
\IEEEoverridecommandlockouts
\usepackage{cite}
\usepackage{amsmath,amssymb,amsfonts}
\usepackage{algorithmic}
\usepackage{graphicx, subfigure}
\usepackage{textcomp}
\usepackage{xcolor}
\usepackage{verbatim}
\usepackage{graphicx}
\usepackage{cite}
\usepackage{booktabs}
\usepackage{placeins}
\usepackage{tikz}
\usepackage{amssymb}
\usepackage{pifont}
\usepackage{tabularx}
\usepackage{multirow}
\usepackage[]{mdframed}
\usepackage{mathtools}
\usepackage{makecell}
\newcommand{\xmark}{\text{\ding{55}}}
\def\BibTeX{{\rm B\kern-.05em{\sc i\kern-.025em b}\kern-.08em
    T\kern-.1667em\lower.7ex\hbox{E}\kern-.125emX}}

\newcommand\InsertBlankPages[1]{% \InsertBlankPages{n} => insert n blank pages
  \foreach \blank in {1,...,#1} {
    \newpage
    \thispagestyle{plain}
    \mbox{}
  }%
}    

\begin{document}
\InsertBlankPages{1}
\begin{center}
\textit{
NOTICE: This paper is the accepted version of our paper with DOI 10.1109/CLOUD62652.2024.00055 that was published to the 17th IEEE International Conference on Cloud Computing
(CLOUD). Please read/cite the final paper/DOI.
©2024 IEEE. Personal use of this material is permitted. Permission from IEEE must
be obtained for all other uses, in any current or future media, including reprinting/republishing
this material for advertising or promotional purposes, creating new collective works, for resale
or redistribution to servers or lists, or reuse of any copyrighted component of this work in other
works.}
\end{center}

\title{Conference Paper Title*\\
{\footnotesize \textsuperscript{*}Note: Sub-titles are not captured in Xplore and
should not be used}
\thanks{Identify applicable funding agency here. If none, delete this.}
}

\title{The State of FaaS: An analysis of public Functions-as-a-Service providers}

\author{\IEEEauthorblockN{Nnamdi Ekwe-Ekwe}
\IEEEauthorblockA{\textit{School of Computer Science} \\
\textit{University of St Andrews}\\
St Andrews, United Kingdom \\
nnee@st-andrews.ac.uk}
\and
\IEEEauthorblockN{Lucas Amos}
\IEEEauthorblockA{\textit{lucasamos.dev} \\
Scotland, United Kingdom \\
contact@lucasamos.dev}
}

\maketitle

\begin{abstract}
Serverless computing is a growing and maturing field that is the focus of much research, industry interest and adoption. Previous works exploring Functions-as-a-Service providers have focused primarily on the most well known providers AWS Lambda, Google Cloud Functions and Microsoft Azure Functions without exploring other providers in similar detail. In this work, we conduct the first detailed review of ten currently publicly available FaaS platforms exploring everything from their history, to their features and pricing to where they sit within the overall public FaaS landscape, before making a number of observations as to the state of the FaaS.
\end{abstract}


\section{Introduction} \label{introduction}

Serverless is an event-driven computing paradigm allowing end-users to write, deploy and run code in the cloud without managing the underlying infrastructure \cite{li2022serverless, wen2023rise}. In the case of Functions-as-a-Service (FaaS), the user writes their function in a specific language, with their code packaged and deployed on the serverless computing platform. The user allocates resources to the function and chooses a method of invocation, with the function executed in response to requests \cite{shafiei2022serverless}. Serverless functions are ephemeral in nature and generally stateless, with the function execution and its underlying resources ceasing to exist post invocation. Serverless billing is typically calculated based on a combination of the function's execution time, provisioned resources and number of invocations \cite{awslambdapricing, googlecloudfunctionspricing, microsoftazurefunctionspricing}. The advantages of serverless computing are becoming increasingly evident. Economically, serverless can allow for significant cost savings with charges incurred based only on when resources are provisioned to run a function \cite{adzic2017serverless}. From an application developer perspective, the developer does not need to manage underlying infrastructure, provision hardware, etc. all they need to do is deploy their function and then process requests \cite{cloudflareserverless}. Surveys from industry show increasing serverless adoption amongst end-users with about half of customers across the largest public cloud providers using at least one serverless solution in their deployments \cite{datadogserverlessreport}.\\

The serverless field is still maturing and as such there is limited literature in addition to several open research areas and challenges \cite{hassan2021survey, li2022serverless, shafiei2022serverless}. In the state of the art, three main public FaaS providers are frequently discussed and analysed. They are AWS Lambda \cite{awslambda}, Google Cloud Functions \cite{googlecloudfunctions} and Microsoft Azure Functions \cite{microsoftazurefunctions}. Studies of serverless adoption and use cases \cite{eismann2020review, pavlov2019serverless} explore important areas such as the most popular languages being used for serverless applications to the most used FaaS platforms, amongst others. However, in recent years there has been an expansion of functionalities offered by several FaaS providers in addition to new FaaS offerings being launched. There has been little to no detailed review of these new FaaS platforms, what their offerings are and how they sit within the overall serverless landscape. Additionally, many new FaaS providers are offering functionality that has deviated from the more traditional Functions-as-a-Service offerings \cite{vercelfunctions, netlifyfunctions, cloudflareserverless} which as of yet, have not been explored in the state of the art.

In this paper, we conduct the first detailed review and analysis of ten publicly available FaaS offerings. We provide an overview of the features offered by each and draw observations from our analysis. The goal of this work is to, (1) provide an overview to the research community and industry practitioners as to the current state of the FaaS landscape and (2), explore their various capabilities in more detail. The structure of this paper is as follows. We start by discussing the methodology we use to obtain the list of the FaaS platforms we review in Section \ref{methodology}. We then discuss the overview and capabilities of each platform (Section \ref{overviewcapabilities}) exploring the history of the platform (Section \ref{history}), the languages supported (\ref{languagessupported}), invocation types (\ref{invocationtypes}), resource configurations (\ref{resourceconfigslimits}), regions supported (\ref{regions}) and pricing models (\ref{pricing}). We then make a number of observations in Section \ref{observations}, discuss related work (Section \ref{relatedwork}) and then conclude this paper in Section \ref{conclusion}.

\section{Methodology} \label{methodology}

Functions-as-a-Service (FaaS) describes self-contained functions hosted on some compute platform that are invoked by some event and carry out some processing \cite{shafiei2022serverless, hassan2021survey, van2017spec, van2018serverless}. To create our collection of FaaS providers to evaluate, we conducted internet searches to garner a list of providers offering public FaaS platforms. Similar to the research state of the art, there appears to be no central up-to-date repository of which providers offer FaaS platforms, with information being highly fragmented across several sources. Our methodology to obtain our list of providers was twofold: 

\begin{enumerate}
    \item \textit{Explore public cloud providers} and research whether or not they offer a FaaS platform. FaaS has grown out of public cloud providers and so we typically see serverless offerings being tied directly to the overall cloud offering. The first known FaaS platform, Lambda, grew out of AWS \cite{lambdalaunch} and so we explored the list of all publicly available cloud providers worldwide to see which offered FaaS platforms to include in our review.
    
    \item \textit{Explore articles/posts written by industry practitioners and developers}: FaaS is being used extensively by industry and as such, many industry practitioners frequently share their experiences/insights using FaaS. For example, Dunelm, a large UK home furnishing retailer is the largest user of Lambda in Europe and frequently writes about its experiences deploying production workloads on Lambda \cite{dunelmserverless, aleios}. Such posts were explored to find public FaaS providers being used by the developer community.
\end{enumerate}

From a combination of these two approaches, our methodology returned a list of \textit{ten} FaaS providers. We found that, aside from the three major cloud companies (Amazon, Google and Microsoft), a number of new FaaS providers have emerged in recent years. We also noted that a FaaS provider, IBM, recently deprecated their FaaS offering - IBM Cloud Functions (which has been cited in prior research in the serverless state of the art such as in work \cite{figiela2018performance}. IBM Cloud Functions was deprecated at the end of 2023 \cite{ibmfunctionsshutdown} and replaced with IBM Code Engine \cite{ibmcodeengine}. We classified our list of providers as \textit{generalised} or \textit{specialised} FaaS offerings. Most of these providers (80\%) are \textit{generalised}, allowing users to upload \textit{any} kind of code written in a number of languages to the platform, with the code executed in response to events. Most of these generalised providers are delivered as part of an overall larger public cloud offering. The other 20\% of those providers are \textit{specialised}, that is, the offering is tightly coupled to deploying more specific kinds of workloads tied to the overall business offering of the provider. The list of providers we explore are shown in Table \ref{tab:serverlessproviders} which are analyse next.

\begin{table}
\caption{FaaS Offerings Evaluated}
\begin{center}
\begin{tabular}{|c|c|c|}
\hline
% \toprule
\textbf{Company} & \textbf{FaaS Offering} & \textbf{Classification}\\
\midrule
\hline
Alibaba & Alibaba Function Compute \cite{alibabafunctioncompute} & Generalised\\\hline
Amazon & AWS Lambda \cite{awslambda} & Generalised\\\hline
Cloudflare & Cloudflare Workers \cite{cloudflareworkers} & Generalised \\\hline
DigitalOcean & Digital Ocean Functions \cite{dofunctions} & Generalised \\\hline
Google & Google Cloud Functions \cite{googlecloudfunctions} & Generalised \\\hline
IBM & IBM Cloud Code Engine \cite{ibmcodeengine} & Generalised \\\hline
Microsoft & Microsoft Azure Functions \cite{microsoftazurefunctions} & Generalised \\\hline
Netlify & Netlify Functions \cite{netlifyfunctions} & Specialised \\\hline
Oracle & Oracle Cloud Functions \cite{oraclefunctions} & Generalised \\\hline
Vercel & Vercel Functions \cite{vercelfunctions} & Specialised \\\hline
\end{tabular}
\end{center}
\label{tab:serverlessproviders}
\end{table}

\section{Overview and Capabilities} \label{overviewcapabilities}

\subsection{History} \label{history}

\begin{figure*}
  \includegraphics[width=\textwidth, scale=1]{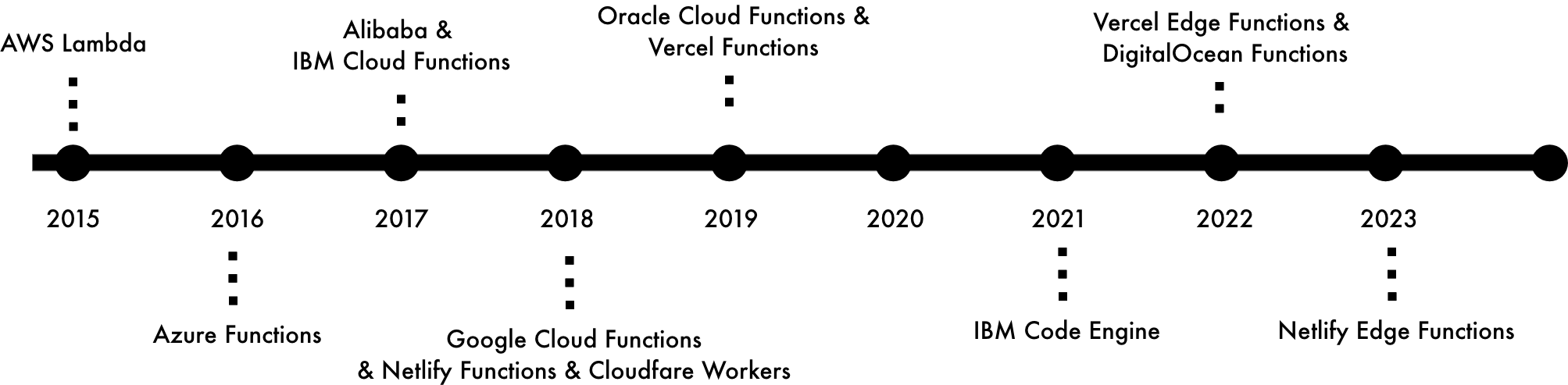}
  \caption{Timeline of General Availability (GA) of FaaS providers}
  \label{fig:faas}
\end{figure*}

The timeline of the creation of the various FaaS providers in Table \ref{tab:serverlessproviders} are shown in Figure \ref{fig:faas}. We garnered this information from a mixture of press releases by the various companies or (where such releases weren't available) from 1) blog posts written by industry practitioners using the FaaS service or 2) the earliest available Release Notes of that FaaS service. AWS provided the first preview of a FaaS offering in 2014 \cite{lambdapreview} before making their platform Generally Available (GA) in 2015 \cite{lambdalaunch}. Microsoft were next, releasing Azure Functions in 2016 \cite{azurefunctionsga}. Google released a preview of Google Cloud Functions in 2016, before their offering became GA two years later \cite{gcpfunctionsga}. Alibaba Functions Compute \cite{alibabaga} and IBM Cloud Functions \cite{ibmcloudfunctionslaunch} were launched in 2017. Cloudflare Workers became GA in 2018 \cite{cloudflareworkersga} with Netlify Functions also released that year \cite{netlifyfunctionspreview}. Oracle Functions \cite{oraclefunctions} was released in 2019. Vercel had a serverless offering in 2019 \cite{vercelfunctions2019}, followed by IBM's new serverless offering Code Engine in 2020 (beta) \cite{codeenginepreview}. This became generally available in 2021 \cite{codeenginega} with IBM Cloud Functions \cite{ibmfunctionsshutdown} being deprecated in 2023 \cite{ibmcodeengine}. Vercel released a new offering, Vercel Edge Functions \cite{vercelfunctionsga} in 2022 with DigitalOcean Functions \cite{digitaloceanfunctionsrelease} also launching that same year. A new offering by Netlify, Netlify Edge Functions \cite{netlifyfunctionsedge} was released in 2023. As mentioned in the previous section, eight out of ten of these providers provide \textit{generalised} serverless compute capabilities. Two of the providers, Vercel and Netlify have FaaS platforms that are more closely tied to the functionality of their overall service. Users of these platforms typically deploy code that is more specifically tied to the overall offering of the company. For example, Vercel is a platform allowing users to deploy web apps and as such, Vercel's Functions are typically used to write functions relating to authentication, image/video processing on web apps, etc. \cite{vercelfunctions}. Netlify is similar to Vercel, also allowing users to deploy web applications, e-commerce platforms, etc. to the web. Their own functions offering is also tied to their web platform offering with functions being used to perform tasks such as updating a database on a form being submitted, sending an email, etc. \cite{earthlyblog}. We next explore the underlying features of each of these platforms.

\begin{table*}[!htbp]
\caption{Languages Supported}
\label{tab:languagessupported}
\begin{center}
\begin{tabularx}{\textwidth}{|X|X|X|X|X|X|X|X|X|X|X|}
\hline
% \toprule
&\textbf{Alibaba} & \textbf{AWS} & \textbf{Azure}& \textbf{Cloudflare}& \textbf{Digital Ocean} & \textbf{Google} & \textbf{IBM} & \textbf{Netlify}& \textbf{Oracle} & \textbf{Vercel}\\
\midrule
\hline
\textbf{Node.js} &\checkmark & \checkmark & \checkmark & \checkmark &\checkmark & \checkmark &\checkmark & \checkmark &\checkmark & \checkmark\\\hline
\textbf{Java} &\checkmark & \checkmark & \checkmark & \xmark &\xmark & \checkmark &\xmark & \xmark &\checkmark & \xmark\\\hline
\textbf{Python} &\checkmark & \checkmark & \checkmark & \xmark &\checkmark & \checkmark &\checkmark & \xmark &\checkmark & \checkmark\\\hline
\textbf{.NET/C\#} &\checkmark & \checkmark & \checkmark & \xmark &\xmark & \checkmark &\xmark & \xmark &\checkmark & \xmark\\\hline
\textbf{Powershell} &\xmark & \checkmark & \checkmark & \xmark &\xmark & \xmark &\xmark & \xmark &\xmark & \xmark\\\hline
\textbf{Ruby} &\xmark & \checkmark & \xmark & \xmark &\xmark & \checkmark &\xmark & \xmark &\checkmark & \checkmark\\\hline
\textbf{Go} &\checkmark & \checkmark & \checkmark & \xmark &\checkmark & \checkmark &\xmark & \checkmark &\checkmark & \checkmark\\\hline
\textbf{PHP} &\checkmark & \xmark & \xmark & \xmark &\checkmark & \checkmark &\xmark & \xmark &\xmark & \xmark\\\hline
\textbf{Custom Runtime} &\checkmark & \checkmark & \checkmark & \checkmark &\xmark & \xmark &\xmark & \xmark &\xmark & \checkmark\\\hline
\end{tabularx}
\end{center}
\end{table*}

\begin{table*}[!htbp]
\caption{Resource Configurations}
\label{tab:resourceconfigurations}
\begin{center}
\begin{tabularx}{\textwidth}{|X|X|X|X|X|X|X|X|X|X|X|}
\hline
% \toprule
&\textbf{Alibaba} & \textbf{AWS} & \textbf{Azure}& \textbf{Cloudflare}& \textbf{Digital Ocean} & \textbf{Google} & \textbf{IBM} & \textbf{Netlify}& \textbf{Oracle} & \textbf{Vercel}\\
\midrule
\hline
\textbf{Maximum Memory} & 32GB & 10GB & 14GB & 128MB & 1GB & 32GiB & 4GB & 1GB & 2GB & 3GB(S), 128MB(E)\\\hline

\textbf{Maximum Temporary Storage} & 10GB & 10GB & 1TB & use-case dependent & NS & NS & NS & NS & NS & NS\\\hline

\textbf{Maximum Timeout} & 24h & 15m & $\infty$ & $\infty$ & 15m & 60m/HTTP, 9m/event-driven & 2m & 10s & 5m & 15m(S), $\infty$(E)\\\hline

\end{tabularx}
\end{center}
\end{table*}

\subsection{Languages supported} \label{languagessupported}

There are a wide variety of languages natively supported by the ten FaaS providers (Table \ref{tab:languagessupported}) with additional languages supported via custom runtimes. AWS natively supports 7 languages in addition to other custom runtimes \cite{awslambdafaqs}. Microsoft \cite{azurelanguagessupported} and Alibaba \cite{alibabalanguagessupported} natively support 6 languages as well as the use of custom runtimes for other languages such as Rust, Go, etc. Google \cite{gcflanguagessupported} and Oracle \cite{ocifunctionslanguagessupported} support 7 and 6 languages respectively but do not support custom runtimes. DigitalOcean \cite{dofunctionslangs} and Vercel \cite{vercelfunctionslangs} support 4 languages natively with DigitalOcean not having custom language support. However Vercel has a custom JavaScript ``Edge Runtime'' available to run lightweight code using a set of WebStandard APIs. IBM \cite{ibmfunctionslangs}, Netlify\cite{netlifyfunctionslangs} and Cloudflare \cite{cloudflareworkerslangs} only support 2 languages. Cloudflare has a custom runtime for languages that compile to WebAssembly, such as Rust, C++, etc.

\subsection{Invocation Types} \label{invocationtypes}

Functions are primarily event-driven with a wide variety of \textit{invocation types} supported. Functions can also be invoked on a schedule as typically seen in a cron job. In Lambda, functions can be invoked by HTTP, Queues, in response to a change in an AWS service or via polling \cite{lambdainvocationtypes}. Multiple triggers to the \textit{same} function are supported - for example a function being invoked by both an HTTP request and change in an S3 bucket, etc. Azure Functions also allow for a wide variety of triggers and bindings \cite{azurefunctionstriggers}. In Azure, a trigger defines how a function is invoked. Unlike Lambda, an Azure function can only have at most \textbf{one} trigger. Bindings allow for Azure functions to be connected to other Azure services and receive and/or send data to other services \cite{azurefunctionstriggers}. Google Cloud Functions split their invocation types into two: (1) HTTP and (2) Event triggers. HTTP triggered functions respond only to HTTP events, whilst functions with Event triggers respond to events such as Pub/Sub, from other Google services (via Eventarc), etc. \cite{gcftriggers}. GCF can only have at most one trigger. Alibaba supports multiple invocation methods (HTTP, via events, etc. \cite{afctriggers}) with similar functionality shown in Oracle \cite{ocfinvocation}. IBM \cite{ibmcodeengineinvocations}, DigitalOcean \cite{dofunctionsinvocation}, Netlify \cite{netlifyfunctionslangs}, Vercel \cite{vercelfunctionsquickstart} and Cloudflare \cite{cloudflaregetstarted} only allow HTTP invocations.

\subsection{Resource configurations and limits} \label{resourceconfigslimits}

We have highlighted the resource configurations common to all providers (Memory, Storage and Timeout) in Table \ref{tab:resourceconfigurations}. Please note that ``Storage'' refers to temporary/ephemeral storage that can be attached to a function, acting as a local disk. Some FaaS providers allow for further configuration parameters that are not common to all of them, such as CPU, concurrency settings, etc. Typically memory size for FaaS is displayed in MB but for readability, we have converted MBs that are $>=1GB$ equivalent to GBs. Please also note that \textit{NS} in the table means not supported as a configurable option. Edge and Standard functions are also denoted as ``E'' and ``S'' in the table. AWS allows the developer to set the memory, timeout and storage size of their Lambda function. The developer can set a maximum memory size and storage of 10GB, and a maximum timeout of 15 minutes \cite{lambdaresourceconfig}, compute capacity scales proportionally with memory and cannot be set independently. The maximum number of simultaneous executions for a function can also be set via concurrency options \cite{lambdaconcurrency}. Alibaba provides more generous maximum memory (32GB) with maximum storage being the same as with Lambda (10GB). The maximum timeout is also the largest of all the FaaS providers at 24h \cite{afclimits}. The number of vCPUs, in addition to the concurrency limit for the function \cite{afcmanage} can also be configured with Alibaba. Azure allows the user to set a maximum function memory of 14GB, with storage being 1TB and an unlimited/unbounded maximum timeout \cite{azurelimits}. It is worth noting that although a function can technically have an unlimited timeout, Microsoft only guarantees this for up to 60 minutes, stating that activities such as OS/runtime/vulnerability patching, etc. can cancel function execution after this time. Also, due to default settings on Azure Load Balancer, HTTP invoked functions have a maximum timeout of 230s \cite{azurelimits}. Google has the most generous maximum memory allocation of all being 32GiB. The maximum timeout for a Google Cloud Function is dependent on its \textit{type} - up to 60 minutes for an HTTP function or 9 minutes for an event-driven function. Similar to Alibaba, Google allows the developer to specify the vCPUs and, like Azure and AWS, the concurrency \cite{gcflimits}. Google, unlike the other aforementioned providers, does not allow the user to add/modify additional storage for a function. IBM CodeEngine has a more limited set of configurations and limits open to the end-user. The end-user's function is limited to running for 2m, with maximum memory size being 4GB. The developer can also specify the ``scale-down'' delay of their function which sets how long an instance remains alive before shutting down - this is at most 6m and is useful to prevent cold start behaviour \cite{codeenginelimits}. DigitalOcean similarly offers a limited set of configuration options. A function on DigitalOcean can only have a maximum memory allocation of 1GB, however a more generous maximum timeout of 15m \cite{dofunctionconfig}. Oracle allows for a maximum memory size of 2GB, no option to specify storage, with a maximum timeout of 5m \cite{oraclefunctionsconfig}.
Cloudflare Workers have a maximum memory of 128MB and have no maximum duration \cite{cloudflarelimits}. The concept of storage for a function is slightly different in Cloudflare Workers, with the developer able to make use of different storage types (key-value stores, object stores, queues, etc.) for their function to persist data \cite{cloudflarestorage}. Cloudflare however does not provide any option to modify temporary/ephemeral storage for a function. Netlify's Functions have a maximum memory of 1GB and maximum timeout of 10s or 15m depending on the type of function (synchronous function invocations have a 10s timeout; a background function more suited to longer-running tasks is up to 15m \cite{netlifybackgroundfunctions, netlifylimits}). Finally, Vercel has two different maximum memory allocations depending on the type of function being run. Standard functions can be configured with memory up to 3GB, with functions using the Edge Runtime capped at 128MB. Similarly, the maximum timeout of a function is different depending on the type of function with Standard functions being up to 15m, and edge functions having an unlimited timeout (suited for streaming data) \cite{vercellimits}.

\begin{figure}
  \includegraphics[scale=0.16]{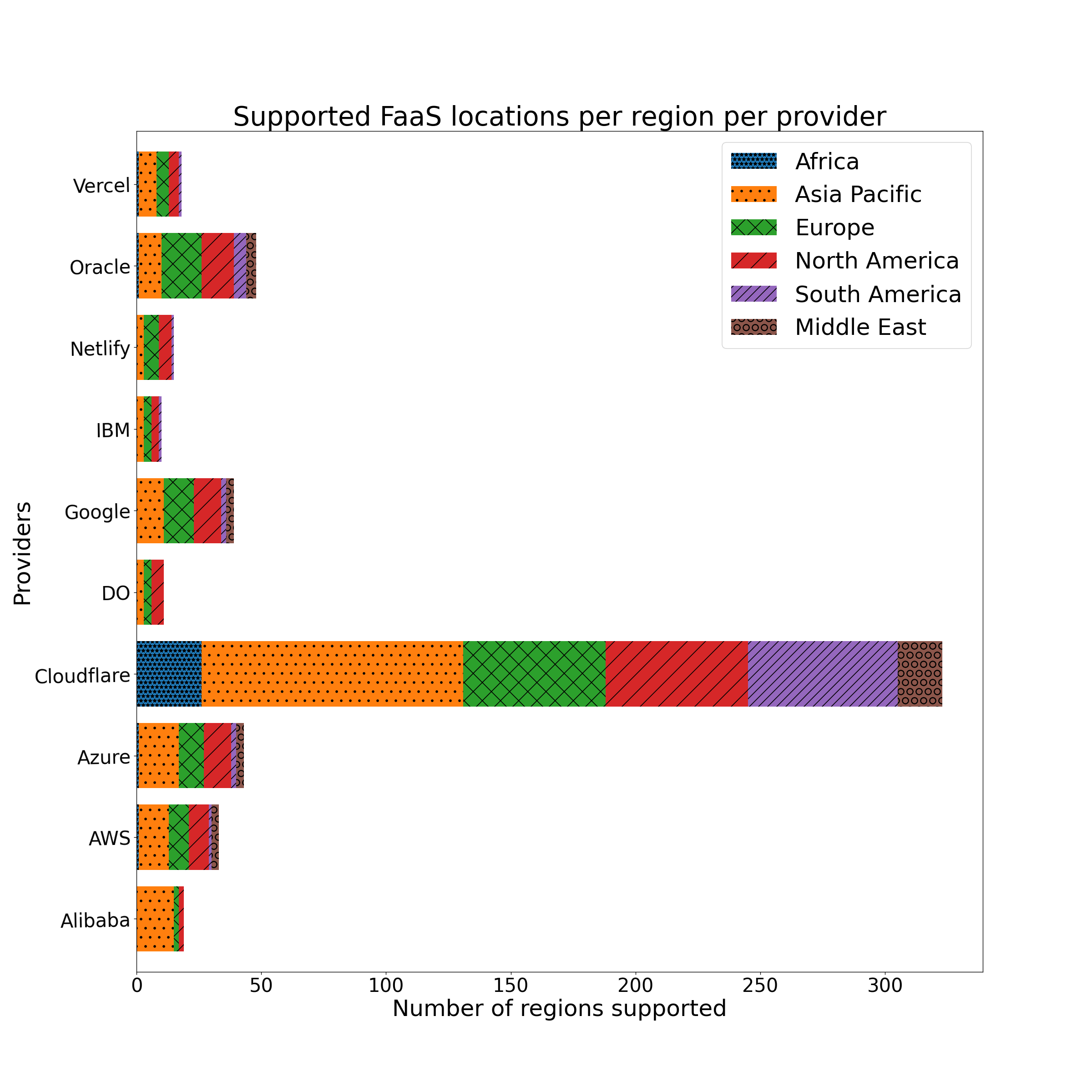}
  \caption{Supported FaaS locations per region per provider}
  \label{fig:regionssuported}
\end{figure}

\subsection{Regions supported} \label{regions}

The various providers support FaaS across their globally distributed locations (Figure \ref{fig:regionssuported}). Alibaba Functions can be deployed in a total of 19 locations \cite{afcregions} with most of them (15) being in the Asia-Pacific region. It is also worth noting that not all Function features are available universally across every supported region - for example GPU-accelerated Function instances. AWS Lambda is available in 33 locations worldwide on every continent with Europe, North America and Asia-Pacific having the most number of locations supported \cite{lambdaregions}. Azure Functions are supported in 43 locations \cite{azureregions}. Cloudflare has the largest number of supported locations (323) with functions deployable worldwide \cite{cloudflareregions}. The reason for this significant coverage is due to Cloudflare being an edge-first provider, and as such covers a wide breadth of locations. DigitalOcean supports one of the fewest number of locations (11), with no locations supported in Africa, South America or the Middle East. Google supports 39 locations \cite{googlefunctionsregions}. IBM supports one less location than DigitalOcean (10) \cite{ibmcodeengineregions}. Netlify supports a subset of AWS's regions as its platform uses AWS as the underlying infrastructure provider \cite{netlifyregions} (15 locations). Oracle supports 48 locations across the world on every continent \cite{oracleregions, oracleregionstwo}. Finally, Vercel supports 18 locations \cite{vercelfunctionsregions}.

\subsection{Pricing} \label{pricing}

The different FaaS platforms have various pricing models. Pricing with FaaS providers is typically made up of two core elements, (1) the number of invocations, and (2) the resources provisioned (CPU/RAM). Other elements such as storage, networking, etc. can differ between providers in the way they are charged (if at all). We report on the core elements that make up the pricing model for each provider next. The common pricing approach amongst FaaS providers is based on (1) the resources consumed and the duration of their use in addition to (2) the number of invocations made. FaaS providers typically charge a fee per GB-second of usage for resources allocated to the function in addition to a flat-fee per $n$ million invocations of a function. The total cost of a function running $P$ can be modelled by Equation \ref{eqn:globalpricing}:
\begin{equation}
\label{eqn:globalpricing}
P = R_{x} + I 
\end{equation}

where $R_{x}$ is the resource consumption charge and $I$ is the invocation charge. For AWS \cite{awslambdapricing, awslambdapricingtwo}, Azure \cite{microsoftazurefunctionspricing} and Oracle \cite{oraclefunctions}, $R_{x}$ is based on the duration of the usage of memory allocated to the function. Google \cite{googlecloudfunctionspricing}, IBM \cite{ibmcodeenginepricing, ibmfreetier}, and Alibaba \cite{alibabapricing} calculate $R_{x}$ based on both the duration of usage of memory as well as CPUs allocated to the function. Finally, with Cloudflare, $R_{x}$ is based solely on duration of CPU usage \cite{cloudflarepricing}. DigitalOcean does not charge an invocation fee, with only an $R_{x}$ charge based on duration of memory used \cite{dofunctionspricing}. The specialised providers have different pricing models from the generalised FaaS providers. Netlify offers plans for their users that come with a fixed amount of invocations per month with the user automatically upgraded to the next tier when usage exceeds that tier \cite{netlifypricing}. Vercel has similar elements to Netlify and Cloudflare, offering plans that include 100GB-hours as well as 500K ``execution units'' for Edge functions \cite{vercelfunctions}. These execution units are equivalent to 50ms of CPU time \cite{vercelexecutionunits}. Two plans are offered - a standard plan and a pro plan with the user allowed to exceed the pro plan GB-hours/execution units allocations for a fee. Enterprise plans are also offered by Vercel for users whose usage does not fit into either of these two plans. Finally, the providers also include a free tier for end users before charging them for function usage (Table \ref{tab:faasfreetier}).

\begin{table}
\caption{FaaS free tier}
\begin{center}
\begin{tabular}{|p{4cm}|p{4cm}|}
\hline
% \toprule
\textbf{Company} & \textbf{Free tier offering}\\
\midrule
\hline
Alibaba & 1M invocations, 400K GB-seconds \\\hline
Amazon & 1M invocations, 400K GB-seconds \\\hline
Cloudflare & 100K invocations per day, 10ms CPU time per invocation\\\hline
DigitalOcean & 90K GiB seconds \\\hline
Google & 400K GB-seconds, 200K GHz-seconds, 5GB network transfer, 2M invocations \\\hline
IBM &  100K vCPU seconds\\\hline
Microsoft & 1M invocations, 400K GB-seconds \\\hline
Netlify & N/A \\\hline
Oracle & 2M invocations, 400K GB-seconds \\\hline
Vercel & ``Hobby'' free tier \\\hline
\end{tabular}
\end{center}
\label{tab:faasfreetier}
\end{table}

\section{Observations} \label{observations}

The ten providers we have analysed offer varying services, features, prices and are available in various locations across the world. We note a number of observations from our analysis:

\begin{mdframed}
\textit{Observation 1}: The FaaS provider landscape has grown, but newer providers are yet to mature to challenge the more established platforms.
\end{mdframed}

Since the turn of the decade (2019/20), we have seen five new FaaS platforms emerge (Figure \ref{fig:faas}). Whilst this adds new competition to the FaaS landscape, the functionality offered by these platforms are yet to mature. We saw this in Table \ref{tab:resourceconfigurations}, with more limited configuration options (such as memory, timeout, etc.) as well as limits on the languages supported (Table \ref{tab:languagessupported}). These limits were also evident in FaaS support across the various regions with DigitalOcean, one of the newest providers, having some the fewest number of locations supported worldwide. In order for these newer FaaS providers to compete with the likes of AWS, Microsoft, Alibaba and Google, more functionality and flexibility needs to be offered.

\begin{mdframed}
\textit{Observation 2}: Whilst the various FaaS providers have many common features/functionalities, some significant differences remain, impacting the types of applications supported.
\end{mdframed}

We noted in our earlier analysis how AWS Lambda allowed for the developer to set multiple triggers simultaneously to the same function (Section \ref{invocationtypes}). There are some use cases where this could be very useful, for example being able to invoke a function externally via an HTTP call whilst also allowing it to be invoked via an internal service that adds a message to a queue. The use-cases and applications for serverless are varied \cite{eismann2021state} and so the developer needs to ensure that their chosen FaaS provider can accommodate their application use-case. We highlighted the functionality by AWS Lambda as it was different to all the other providers. These providers either (1) limited the user to choosing at most one invocation type for their function or (2) had a limited set of invocation options supported by default. This was also seen in the case of languages supported with some providers supporting a large number of languages by default and allowing for other languages to be implemented via custom runtimes. These various constraints need to be carefully considered by the application developer when choosing a particular FaaS provider.

\begin{mdframed}
\textit{Observation 3}: Some of the FaaS providers are making the edge available for developers to deploy workloads.
\end{mdframed}

Running functions on the network edge, reducing latency and speeding up function serving and handling of requests is now being introduced across several platforms such as AWS, Netlify, Vercel and Cloudflare. The advantages of the edge with its low latency and how this can be leveraged for functions, handling requests close to the users in geographically distributed locations is clear. Functionality that is run on these providers is very web-oriented with functions on Netlify and Vercel typically tied to their web application workloads such as authentication, form processing, etc. Cloudflare is more generalised in not restricting the kinds of workloads that can be run on the edge. From our work, we note how edge locations attract significantly reduced resources (128MB maximum for Vercel and Cloudflare) and so decisions on which functions to run on the edge needs to be carefully considered.

\begin{mdframed}
\textit{Observation 4}: JavaScript, Python and Go are the languages most frequently supported by the FaaS providers
\end{mdframed}

JavaScript/Node.js is the only language supported by all the FaaS platforms. This was closely followed by Python and Go that were both tied on 8 providers. Java and C\# were both tied on 5, with the other languages being sparsely supported by the platforms. The use of the custom runtimes was supported by only 4 out of the 10 providers. Custom runtimes are certainly advantageous in accommodating for any gaps in native language support but this feature is not well established across the explored FaaS providers. It would be interesting to see the performance of these custom runtimes across the providers that support them and if there are any differences. Works \cite{cordingly2020implications, jackson2018investigation} have explored how language affects function performance and price.

\begin{mdframed}
\textit{Observation 5}: Resources and configuration options are generally generous amongst providers, with the newer FaaS providers providing more limited resources.
\end{mdframed}

Most of the providers offer generous resource configurations for their users' function workloads. Alibaba overall has the most generous, offering the largest amount of memory and timeout of all the providers (Table \ref{tab:resourceconfigurations}). Google offers a slightly higher memory offering than Alibaba but does not provide any separate storage configuration and has a significantly smaller timeout. Azure offers the highest amount of storage and a high amount of memory per function with a function timeout that is not limited (at least 60m uninterrupted). Cloudflare does allow for an unlimited timeout albeit with a relatively small amount of memory available. The newer providers (DigitalOcean, IBM, Oracle) have the most limited set of configurations available of all the generalised providers. Vercel is quite generous with its function configuration (depending on whether the function is standard or edge) however those functions are tied directly to its overall web offering with Netlify also being quite limited.

\begin{mdframed}
\textit{Observation 6}: There are differences amongst the various providers' pricing models, but core elements are common to them all.
\end{mdframed}

In general, pricing follows a duration/resource model, whereby an end-user is charged for the duration of their function and the resources they consume. They are also additionally charged for the number of invocations of that function. Cloudflare is the only provider that bills based on CPU-time rather than duration, a move they argue is advantageous as it does not include ``idle time'' of the function waiting on network requests/IO \cite{cloudflareadvantage}. There is work exploring optimising FaaS function placement to take advantage of price \cite{elgamal2018costless, mahajan2019optimal} however an exploration of these various differences using real-world workloads across the providers reviewed would further add to exploring this area.

\section{Related Work} \label{relatedwork}

Several related works exploring the state of the art have been produced by both the research and industry communities. Key surveys \cite{li2022serverless, shafiei2022serverless, eismann2021state, wen2023rise}, discuss the state of the art of serverless, exploring everything from the evolution and creation of serverless, to common use cases, to the key challenges and opportunities in the field, etc. From industry, we note works such as DataDog's The State of Serverless report \cite{datadogserverlessreport} also exploring key trends/statistics within Serverless today. Empirical studies understanding and comparing multiple FaaS platforms \cite{figiela2018performance, chowhan2018hands, perez2018cost, malawski2018benchmarking} have been produced primarily comparing the big three (Azure, AWS and Google) in addition to IBM OpenWhisk. We also note works from industry such as those by Mikhail Shilkov \cite{mikhail1} exploring a wide range of topics, such as exploring cold start behaviour across the various platforms, to testing their scalability, etc. Several severless benchmark frameworks \cite{maissen2020faasdom, yu2020characterizing, copik2021sebs} have also garnered empirical evidence on the performance and cost of executing functions on the big three FaaS providers, drawing conclusions from their experiments.

\section{Conclusion} \label{conclusion}

In this paper, we have provided the first detailed review of all ten currently available public FaaS providers. We have discussed various elements of each provider from their history, invocation types, features/resource configurations supported, regions covered, and pricing models employed. We have also made a number of observations from our research as to the state of public FaaS providers and provided insight as to where further work can be performed. Empirical evaluations of various facets of these additional providers would be advantageous to add to already burgeoning research in this area.

\section{Acknowledgements} \label{acknowledgements}

We want to thank the anonymous reviewers for their time and valuable feedback.This study is based on the use of primary sources accessed from several archives. A full list of the sources is available in the reference list.

\bibliographystyle{IEEEtran}
\bibliography{references.bib}

\end{document}